\documentclass[12pt]{article}

\usepackage{natbib}

\usepackage{bbm}

\usepackage{algorithm}

\usepackage{inputenc,crop,graphicx,amsmath,array,color,amssymb,flushend,stfloats,amsthm,chngpage,times,fullpage}
\usepackage[LY1]{fontenc}

\begin{document}

\title{Extrema-weighted feature extraction for functional data} 
\author{Willem van den Boom\,$^{\text{1}}$ \and Callie Mao\,$^{\text{1}}$ \and Rebecca A.\ Schroeder\,$^{\text{2}}$ \and David B.\ Dunson\,$^{\text{1}}$ (\texttt{dunson@duke.edu})}
\date{$^{\text{1}}$Department of Statistical Science, Duke University, Durham, NC, 27708, US and \\
$^{\text{2}}$Department of Anesthesiology, Duke University School of Medicine, Durham, NC, 27710, US. \\[2ex]
\today}

\maketitle

\abstract{
\noindent
\textbf{Motivation:}
Although there is a rich literature on methods for assessing the impact of functional predictors, the focus has been on approaches for dimension reduction that can fail dramatically in certain applications.
Examples of standard approaches include functional linear models, functional principal components regression, and cluster-based approaches, such as latent trajectory analysis.
This article is motivated by applications in which the dynamics in a predictor, across times when the value is relatively extreme, are particularly informative about the response.
For example, physicians are interested in relating the dynamics of blood pressure changes during surgery to post-surgery adverse outcomes, and it is thought that the dynamics are more important when blood pressure is significantly elevated or lowered.\\
\textbf{Methods:}
We propose a novel class of extrema-weighted feature (XWF) extraction models.  Key components in defining XWFs include the marginal density of the predictor, a function up-weighting values at high quantiles of this marginal, and functionals characterizing local dynamics.
Algorithms are proposed for fitting of XWF-based regression and classification models, and are compared with current methods for functional predictors in simulations and a blood pressure during surgery application.\\
\textbf{Results:}
XWFs find features of intraoperative blood pressure trajectories that are predictive of postoperative mortality.
By their nature, most of these features cannot be found by previous methods.

\section{Introduction}

Functional predictors are ubiquitous in applications. Think for instance of images or time-series data. One often wants to use these rich data to predict outcomes or classify subjects into groups. For example, one might want to predict returns based on the stock index trajectory over time or label the contents in a picture. Even though the picture and the stock index are both functional data, their nature is clearly different, showing the diversity in functional data analysis (FDA) applications, and consequently methods.

This paper presents a new FDA method motivated by clinical questions regarding blood pressure measurements. 
Intraoperatively, blood pressure is measured either continuously via invasive monitors or at intervals of 30 seconds to 3 minutes.
See Figure~\ref{BPtraj} for example trajectories.
\begin{figure}[tb]
	\centering
	\includegraphics[width=3.2in]{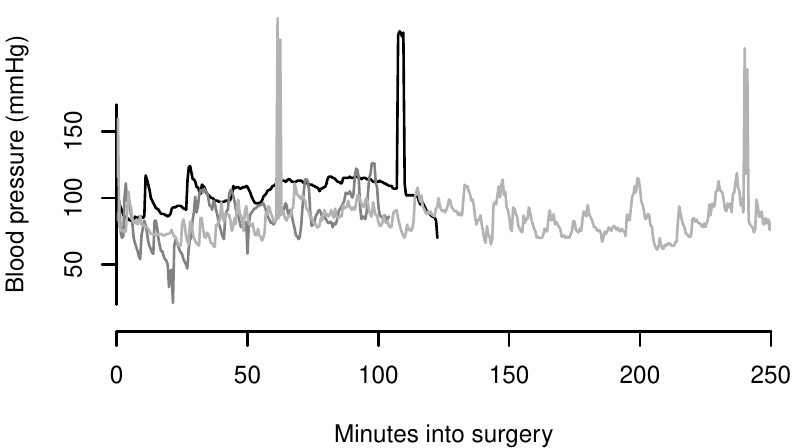}
	\caption{The mean arterial blood pressure measurements of three surgeries during their first three hours.}
	\label{BPtraj}
\end{figure}
The effects of various blood pressure levels are extensively studied. Nonetheless, questions from clinicians remain, especially regarding the relevance of various degrees of blood pressure variability \citep{Mancia2000}.

Variability has been established as an important biomarker~\citep{Parati2013}, even though often not fully considered. For instance, it is suggested that high average blood pressure can be a result of blood pressure variability such that established associations between high blood pressure and certain outcomes might be driven by variability. For instance, \citet{Rothwell2010} found that blood pressure variability can predict stroke better than mean blood pressure, and that the effect of variability interacts with the mean.
\citet{Sternbach2002} show that anesthesiologists have a big influence on perioperative blood pressure variability.

It is not clear what features of the dynamic variations in blood pressure are most relevant \citep{Aronson2010}.
For example, anesthesiologists want to identify those aspects of the dynamics that
predict adverse postoperative outcomes, such as postoperative mortality.
The main goal is to identify, in real time, those patients
that are experiencing adverse blood pressure dynamics.  Anesthesiologists can then intervene and modify the
dynamics to hopefully prevent subsequent organ injury.
It is thought that abrupt changes in blood pressure, especially when associated with severe systolic hypertension and shear stress, significantly increase this risk \citep{Aronson2011}.
However, this is mostly based on poor quality evidence and there is no clear idea what types of dynamics are most or least harmful.

Previous studies have used simple metrics of blood pressure variability, such as the standard deviation of measured values, which ignores the order in which the measurements were obtained, and average real variability \citep{Hansen2010, Mena2005}. These features capture only a very specific type of blood pressure dynamic. Moreover, they do not allow for blood pressure values at the extremes to be treated differently from those in the normal blood pressure ranges.

Trying to relate functions, here blood pressure trajectories, with an outcome, such as postoperative mortality, is often done in FDA. See, for instance, \citet{Ramsay2005}. Usually, one extracts features from the functions, which are then fed into a regression or classification model. Popular methods for feature extraction are functional principal component analysis \citep{Jones1992} or basis expansions, which include wavelets \citep{Morris2003,Morris2008} and Fourier transforms. \citet{Giraldo2010} describe feature extraction jointly with dimension reduction to motivate a weighting of features extracted from functional data. Despite the large literature, existing functional data analysis methods are clearly unsatisfactorary for our motivating blood pressure application, as well as for other related applications.

Most existing approaches assume that the various functions share the same domain, while the time and progression of surgery varies substantially across patients, even when stratifying by type of procedure.
One could naively normalize the lengths of surgeries, but such standardization removes valuable information on time and has no clinical justification.
The reason to place the curves on a common time domain is that usual functional data analysis methods focus on characterizing shared versus individual-specific components of variability in this trajectory specific to each time.  If there are similarities in the dynamics over time (e.g., a spike upwards in the curve), then the curves need to be aligned so that the dynamic features (e.g.\ the spike) occur at the same time to enable detection. For the blood pressure data, such time alignment is artificial; instead, we want to characterize similarities and differences in the dynamics in the trajectories even if these dynamics occur at varying times during surgeries of very different lengths.

These shortcomings of existing methods for this application and the interest of clinicians in local effects,
with the extremes of particular importance, provide motivation for this article.  Our focus is on proposing a new approach for feature extraction for functional predictors that address these issues.
Its main building blocks are a flexible definition of variability, or more generally local effects, and a framework that allows for extremes to contribute more than average values. 
Functions measuring local effects, or variability, are averaged over the domain of the functional predictor. This is a weighted average where the weight is influenced at each point in the domain by how extreme the functional predictor is there. This average for a specific local-effect function is referred to as an extrema-weighted feature (XWF).

\subsection{The blood pressure data and the outcome} \label{application}

We have access to the blood pressure data from 80,065 surgeries performed at the Duke University Health System between 2000 and 2014. We consider the mean arterial pressure (MAP). All MAP measurements below 10 and above 250 are discarded as measurement errors. Most measurements are 30 seconds apart but not all. The methods presented can deal with unevenly space measurements but might behave unexpectedly in the presence of large gaps in measurements. Therefore, cases with gaps of more than 5 minutes in their blood pressure measurements are removed, also because they potentially involve unusual circumstances. Additionally, we remove blood pressure trajectories that are shorter than 30 minutes.

The outcome of interest is 30-day postoperative mortality. The medical database includes mortality records which indicate whether someone died within 30 days after surgery. Clinically, certain postoperative complications, such as infection or reduced renal function, are more interesting to be related with intraoperative blood pressures. However, mortality data are more reliable and have less missing data issues than other outcomes, and hence provide a clearer application to test our new method on. 

To obtain a more homogeneous population of procedures,
we restrict ourselves to non-cardiac surgeries. 
Other features considered are gender, age, and body mass index (BMI). Any cases with these missing, or a BMI above 400, most likely a data error, are excluded.
This leaves us with 18,657 surgeries. Figure~\ref{BPtraj} shows a couple of these blood pressure trajectories.

\section{Methods}

This section describes the proposed extrema-weighted feature extraction as well as two alternative approaches we will use for comparison.

\subsection{Notation and proposed model} \label{model}

Let $y_i$ denote the response for subject $i$, for $i=1,\ldots,n$.  Let $x_i:\mathcal{T}_i \to \Re$, where $\mathcal{T}_i \subset \mathcal{T} \subset \Re$, denote a functional predictor for subject $i$, here the MAP trajectory, with $\mathcal{T}_i$ the time domain across which the predictor is observed for that subject.
We restrict ourselves for simplicity to functions that are curves over time, though extensions to functions in more than one dimension are possible.
Let $z_i = (z_{i1},\dots,z_{iq})^T$ denote $q$ other covariates that may be important in predicting $y_i$. For the blood pressure application, we consider gender, age, BMI, and duration of surgery such that $q=4$.

We assume dense measurements of $x_i(t)$ are available at successive locations $t_i = (t_{i1},\ldots,t_{in_i})^T$ within $\mathcal{T}_i$.  In the presence of short blocks of time with intermittent missingness within $\mathcal{T}_i$, we use linear interpolation to fill in the gaps.  Although we could estimate our features without this interpolation, simply using the times that are available for a patient, we wish to account for the possibility that the observed trajectories at adjacent times to the missing data block are informative about the occurrence of missing values. One can alternatively use more advanced interpolation or model the measurements as noisy observations of the underlying function $x_i$, though we do not consider such approaches in this article.

We let $f$ denote the marginal density of $x_i(t)$. In the application, $f$ is the marginal density of blood pressure measurements.
We define this marginal density to satisfy $x_i(t) \sim f$, with $t$ sampled uniformly from $\mathcal{T}_i$, and the density $f$ defined across the population of subjects.
In addition, let $F$ denote the cumulative distribution function (CDF) of $f$.
Motivated by the application in Section~\ref{application}, we want to define a method for upweighting the importance of local features of $x_i$ for times $t$ where $x_i(t)$ is in the tails of $f$.  In the application, these tails correspond to unusually high or low blood pressure values calibrated relative to the distribution of blood pressure values across individuals having non-cardiac surgical procedures.  Let $u_i(t) = F\{ x_i(t) \}$, for $t \in \mathcal{T}_i$, denote trajectories that are transformed to have standard uniform marginal density.

\subsubsection{Local features}

Let $\psi_j(x_i, t)$, for $j=1,\ldots,p$, be functions that characterize different aspects of the local dynamics of the trajectory $x_i$ at time $t$. Let us define $(c)_+ = \max(0, c)$ and $(c)_- = \max(0, -c)$.  In this paper, $p=4$, $\psi_1(x_i, t) = 1$, $\psi_2(x_i, t) = x_i(t)$ as the level of the trajectory at that time point, $\psi_3(x_i, t) = \left( x_i'(t) \right)_+$ as the rate of increase, and $\psi_4(x_i, t) = \left( x_i'(t) \right)_-$ as the rate of decrease.

\subsubsection{Extrema weighting}

Weight functions $\omega_j(u)$ on $[0,1]$ pair with the functions $\psi_j$. They allow differential weights of the local functional 
dynamic features according to the quantile $u_i(t)$. This is motivated by the hypothesis that a specific blood pressure local functional feature affects the outcome in varying degrees depending on whether it appears at low, medium, or high blood pressures.

Due to the difference in nature of abnormally high versus low blood pressures, we split $\omega$ into two parts; one upweighting low ($\omega_\text{L}$) and one upweighting high ($\omega_\text{R}$) blood pressures. Specifically, we choose
\[
	\omega_j(u) = \omega_{j\text{L}}(u) + \omega_{j\text{R}}(u)
\]
with
\[
	\omega_{j\text{L}}(u) = \begin{cases}
		1, &u \leq b_{j\text{L}} \\
		\frac{1-u}{1-b_{j\text{L}}}, &u > b_{j\text{L}}
	\end{cases}
\]
and
\[
	\omega_{j\text{R}}(u) = \begin{cases}
		1, &u \geq b_{j\text{R}} \\
		\frac{u}{b_{j\text{R}}}, &u < b_{j\text{R}}
	\end{cases}.
\]
Here, $b_{j\text{L}}\in(0,1/2]$ and $b_{j\text{R}}\in[1/2,1)$ are parameters controlling the weighting near zero and one. Figure~\ref{weightFunctions} contains examples.
\begin{figure}[tb]
	\centering
	\includegraphics{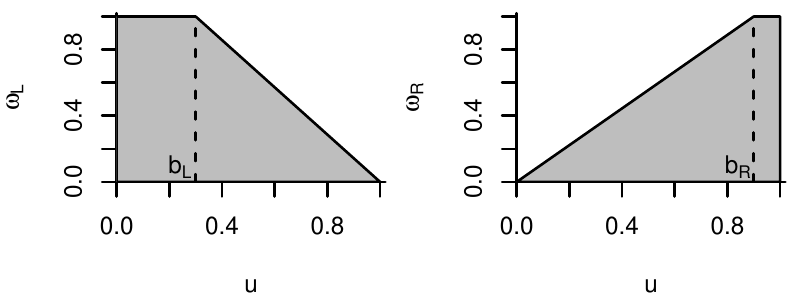}
	\caption{Examples of $\omega_{j\text{L}}$ and $\omega_{j\text{R}}$ with $b_{j\text{L}} = 0.3$ and $b_{j\text{R}} = 0.9$, respectively.}
	\label{weightFunctions}
\end{figure}
Weight functions that are more concentrated near zero or one will weight values in the further extremes of the tail more highly.  We will estimate $b_{j\text{L}}$ and $b_{j\text{R}}$ based on the data.

\subsubsection{Extrema-weighted features}
Employing the aforementioned components, we define the extrema-weighted features (XWFs) as 
\begin{equation}
\begin{aligned}
w_{ij}
&= \frac{1}{T_i} \int_{\mathcal{T}_i} \omega_j\{ u_i(t) \} \psi_j ( x_i, t ) dt \\
&= \frac{1}{T_i} \int_{\mathcal{T}_i} \omega_j[ F\{x_i(t)\} ] \psi_j ( x_i, t ) dt \\
&= \underbrace{\frac{1}{T_i} \int_{\mathcal{T}_i} \omega_{j\text{L}}\{ u_i(t) \} \psi_j ( x_i, t ) dt}_{w_{\text{L} ij}}
+ \underbrace{\frac{1}{T_i} \int_{\mathcal{T}_i} \omega_{j\text{R}}\{ u_i(t) \} \psi_j ( x_i, t ) dt}_{w_{\text{R} ij}}
\label{eq:xwf}
\end{aligned}
\end{equation}
where $T_i = |\mathcal{T}_i|$ is the total time across which the trajectory is observed for individual $i$. Since this time may be informative about the outcome $y_i$, we included $T_i$ as one of the covariates in $z_i$.

Finally, the response $y_i$ is modeled as a generalized additive model (GAM) \citep{Hastie1986, Hastie1990} with predictors $x_i$, in terms of $w_{\text{L} i}$, and $w_{\text{R} i}$, and $z_i$:
\begin{eqnarray} \label{lm}
y_i \sim g\!\left(
	\alpha + \sum_{j=1}^q \gamma_j z_{ij} + \sum_{j = 1}^p\beta_{\text{L}j}\!\left(w_{\text{L} ij}\right) + \sum_{j = 1}^p\beta_{\text{jR}}\!\left(w_{\text{R} ij}\right)
\right)
\label{eq:y}
\end{eqnarray}
with $\alpha$ an intercept, $\gamma_1,\ldots,\gamma_q$ coefficients for the covariates $z_i$, and $\beta_{\text{L} 1}, \beta_{\text{R} 1},\ldots,\beta_{\text{L} p},\beta_{\text{R} p}$ (non-linear) functions on the XWFs.
In the blood pressure application, we use logistic regression. That means that $g(s)$ above is a Bernoulli distribution where the success probability is given by the logistic function, $\frac{1}{1+\exp(-s)}$.

\subsubsection{Estimation Algorithm}

We estimate the density $f$ through a weighted kernel smoothing approach.
In particular, we treat the data as arising from a stratified sampling design, and assign each blood pressure measurement $x_i(t_{ik})$, $k = 1,\dots,n_i$, sampling weight $1/T_i$, which is inversely proportional to the surgery duration.
From an estimate of the density $f$, we can obtain an estimate of the CDF $F$.

The integral in defining the XWFs in \eqref{eq:xwf} is one-dimensional and can be accurately approximated using a variety of numerical methods.  Then, once accurate approximations of the feature vectors $w_{\text{L}i}$ and $w_{\text{R}i}$ are produced, conditionally on the unknown $b_\text{L}$ and $b_\text{R}$ in the weight functions $\omega$, we can easily obtain maximum likelihood estimates for the coefficients in \eqref{eq:y}. Therefore, a grid search can find the $b_\text{L}$ and $b_\text{R}$ that maximize the GAM likelihood.
\begin{algorithm}[tb]
	\caption{Adaptive grid search to optimize the weighting functions} \label{adaptiveSearch}
	
	\textbf{Initialization:} Set $b_{j\text{L}} = 0.25$ and $b_{j\text{R}} = 0.75$ for $j = 1,\dots,p$.
	\vspace*{\baselineskip}
	
	For $l=1,\dots,L$, do the following for $j = 1,\dots,p$:
	\begin{enumerate}
	\item
	Compute the likelihood both for when $b_{j\text{L}}$ is $2^{-1-l}$ larger than its current value, equal to its current value, and when it is $2^{-1-l}$ smaller than its current value, while not changing any other parameters. Set $b_{j\text{L}}$ equal to the value out of these three that yielded the largest likelihood.
	\item
	Compute the likelihood both for when $b_{j\text{R}}$ is $2^{-1-l}$ larger than its current value, equal to its current value, and when it is $2^{-1-l}$ smaller than its current value, while not changing any other parameters. Set $b_{j\text{R}}$ equal to its value out of these three that yielded the largest likelihood.
\end{enumerate}

	\textbf{Return:} Values for $b_{j\text{L}}$ and $b_{j\text{R}}$, $j=1,\dots,p$.
\end{algorithm}
We employ an adaptive grid search akin to binary search, detailed in Algorithm~\ref{adaptiveSearch}.
We use $L = 3$ in this paper such that we get $b_{j\text{L}}$s and $b_{j\text{R}}$s at a resolution of $2^{-4}$.

\subsection{Alternative approaches for comparison}

To evaluate the extrema-weighted feature extraction in perspective, we consider two other approaches. A measure from the medical literature, average real variability, and a method based on the power spectrum of blood pressure trajectories which is closer to traditional FDA approaches.

\subsubsection{Average real variability}

\citet{Mena2005} proposed average real variability (ARV) as an improvement over standard deviation as a measure of blood pressure variability.
ARV has been used to measure the effect of variability on health outcomes \citep{Hansen2010}.
We consider the version from \citet{Hansen2010} that takes into account differences in times between subsequent measurements. That is
\[
	\text{ARV}_i = \frac{1}{t_{in_i} - t_{i1}}
	\sum_{k = 2}^{n_i} \left( t_{ik} - t_{i(k-1)} \right) \left| x_i(t_{ik})  - x_i(t_{i(k-1)}) \right|
\]
in our notation.

To compare the extrema-weighted features with this more straight-forward measure, we fit the data in a generalized additive model with one of the predictors being the ARV. Specifically, the alternative approach consists of replacing the $2p$ extrema-weighted features in \eqref{lm} by the ARV.

\subsubsection{Power spectrum}

Being able to capture local dynamics rather than global effects is an important characteristic of extrema-weighted features. An issue with global effects is that they require some kind of alignment across observations of the functional predictor. By transforming the functional predictor to the frequency domain, one also moves away from this alignment requirement. Therefore, we consider a standard FDA approach on the power spectrum of the functional predictor.

We compute the power spectrum of each $x_i$ via a fast Fourier transform. Since a spectrum is a function itself, we treat it as a functional predictor. Specifically, we compute its functional supervised principal components \citep{Bair2004, Bair2006} and replace the extrema-weighted features in \eqref{lm} by the 3 largest principal components. That is, functional supervised principal component analysis on the power spectrum of the functional predictor fed into a generalized additive model. This represents a more traditional FDA approach to the type of data considered.

\subsection{Inference}

Our extrema-weighted features algorithm estimates the weighting functions by maximum likelihood, which are then fed into the generalized additive model. The $p$-values that the GAM provides however do not take into account any estimation uncertainty in the weighting functions. To obtain principled measures of statistical significance, we obtain $p$-values via a randomization test.

We reorder the elements in the outcome vector $y$ at random and fit the extrema-weighted features model to the resulting data and record the reported $p$-values. By construction, any association between the predictors and the outcome $y$ in these data is by chance. Also, the recorded $p$-values are sample statistics. By repeating the above, we can estimate the sampling distribution of these $p$-values if there is no relation between the predictors and the outcome. By computing what proportion of the time the $p$-values from the original data are less than those from the randomized data, we obtain a measure of statistical significance. We call this proportion the $p$-value for the remainder of the paper.

For consistency, the $p$-values for the alternative approaches, the ARV and the power spectrum, are obtained in the same fashion. An added benefit of this approach is that it does not require assumptions on the data generating process for the inference to be valid.

\section{Results} \label{results}

This section starts with two simulation studies to understand the performance and interpretation of all three methods considered. It concludes with the results from the blood pressure application.

\subsection{Frequency simulation}
\label{frequencySim}

Let $n = 1,000$, and, for $i=1,\dots,n$; $n_i=500$ and $t_{ik}=k$ for $k = 1,\dots,n_i$. We draw $\phi_i$ and $m_i$ from a standard uniform distribution independently for $i=1,\dots,n$. The functional predictors are defined by
\[
	x_i(t) = \sin\!\left(\frac{\pi\phi_i}{25}t\right) + 10m_i,
	\qquad
	i = 1,\dots,n.
\]
Let $q = 2$ and sample the matrix entries of $z$ i.i.d.\ from a standard normal distribution.
Define $\theta_i = 20\phi_i\min(10m_i,7)$. Then, the outcome $y\in\{0, 1\}$ is sampled according to
\[
	P(y_i=1) = \frac{1}{1+\exp\!\left( \mathrm{median}\!\left(\{\theta_j\}_{j=1}^n \right) - \theta_i \right)},
\]
independently for $i=1,\dots,n$.

\begin{table}[tb]
\centering
\caption{Results from the frequency simulation\label{tabSim1}} {\begin{tabular}{@{}lll@{}}
\hline
Method & Parameter & $p$-value \\
	\hline
	
	Extrema-weighted features & $\beta_{\text{L} 1}$& .25 \\
	& $\beta_{\text{R} 1}$& .33 \\
	& $\beta_{\text{L} 2}$& .39 \\
	& $\beta_{\text{R} 2}$& > .99 \\
	& $\beta_{\text{L} 3}$& .57 \\
	& $\beta_{\text{R} 3}$& <.01 \\
	& $\beta_{\text{L} 4}$& .45 \\
	& $\beta_{\text{R} 4}$& .35 \\
	\hline
	
	Average real variability & ARV & < .01 \\
	\hline
	
	Power spectrum & PC 1 & < .01 \\
	& PC 2 & .37 \\
	& PC 3 & < .01 \\
	\hline
\end{tabular}}{}
\end{table}
Table~\ref{tabSim1} contains the results for the extrema-weighted features, ARV, and the power spectrum. The $p$-values for $z$ were not significant and were left out.
Figure~\ref{sim1PCs} contains the principal components used in the power spectrum method.
\begin{figure}[tb]
	\centering
	\includegraphics{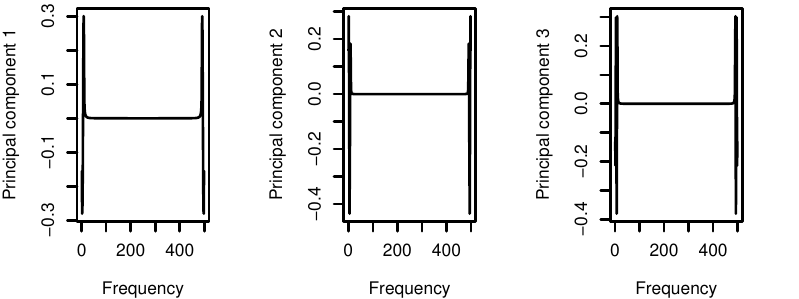}
	\caption{Top three functional supervised principal components of the power spectra from the frequency simulated data.}
	\label{sim1PCs}
\end{figure}
The weighting function of the statistically significant extrema-weighted feature (with $p$ < .05) was estimated as $b_{\text{R}3} = .66$.
The corresponding smooth GAM function is shown in Figure~\ref{GAMfunctionsFreq}.
\begin{figure}[tb]
	\centering
	\includegraphics{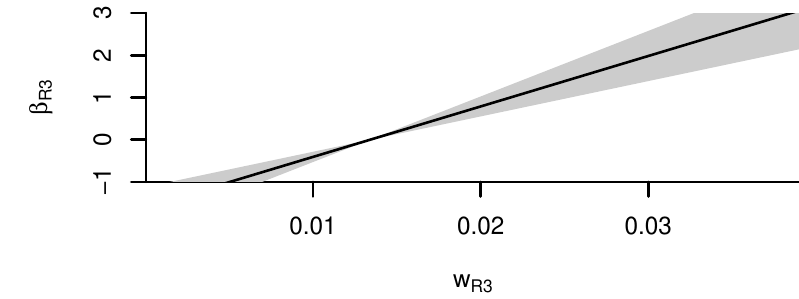}
	\caption{Statistically significant smooth function from the generalized additive model for the frequency simulation with 95\% confidence region in gray.}
	\label{GAMfunctionsFreq}
\end{figure}
The confidence region is from the GAM and ignores uncertainty in our estimation of the weighting functions.

Since ARV increases with $\phi_i$, ARV finds a trend between variability and the outcome in the simulated data. Similarly, $\phi_i$ directly relates to the frequencies in the power spectrum. Therefore, the power spectrum method also finds an effect. However, it is hard to interpret what the various spectrum principal components mean, with all three of them being rather similar with loadings at the extremes of the spectrum.

Extrema-weighted feature selection picks up that an increased derivative, which is proportional to $\phi_i$, corresponds to a higher $P(y_i=1)$. Additionally, this effect is weighted, in that it is more prominent for higher values of $x_i(t)$. This corresponds with the factor $\min(10m_i,7)$ in $\theta_i$.
The results from the XWF-based analysis are consistent with the underlying data-generating process, and richer and clearer in interpretation than the alternative methods.

\subsection{Autoregressive simulation}

The matrix $z$, $n$, $n_i$, $t$, and $\phi$ are the same as in the frequency simulation from the previous section. We draw the marginal variances $v_i$, $i = 1,\dots,n$, i.i.d.\ from a uniform distribution on $[1, 10]$ and the measurements of $x_i$ as follows: Sample $x_i(1)$ from $\mathcal{N}(0, v_i)$ and
\[
	x_i(k) \sim \mathcal{N}\left(\phi_i x_i(k-1), \left(1 - \phi_i^2\right)v_i \right),
	\qquad
	k = 2,\dots,n_i,
\]
independently for $i = 1,\dots,n$.
The outcome $y$ is sampled from the distribution $P(y_i=0)=1-p_i$ and $P(y_i=1)=p_i$ independently, for $i=1,\dots,n$, with
\[
	p_i = 0.01 + 0.99 \cdot \mathbbm{1}\!\left[\phi_i > 0.2 \right] \frac{w_i}{\max_j w_j}
\]
where $\mathbbm{1}\!\left[\cdot\right]$ is the indicator function and
$w_i = \sum_{k=1}^{n_i} \mathbbm{1}\left[ x_i(k) > 2 \right]$.
That is, a persistence above $\phi_i > 0.2$ corresponds with an increased $P(y_i=1)$ where the magnitude of the increase is weighted by the proportion of time the functional predictor $x_i$ is greater than 2.

\begin{table}[tb]
\centering
\caption{Results from the autoregressive simulation with weighting\label{tabSim2}} {\begin{tabular}{@{}lll@{}}
\hline
Method & Parameter & $p$-value \\
	\hline
	
	Extrame-weighted features & $\beta_{\text{L} 1}$& < .01 \\
	& $\beta_{\text{R} 1}$& .26 \\
	& $\beta_{\text{L} 2}$& .35 \\
	& $\beta_{\text{R} 2}$& .38 \\
	& $\beta_{\text{L} 3}$& .90 \\
	& $\beta_{\text{R} 3}$& .28 \\
	& $\beta_{\text{L} 4}$& .19 \\
	& $\beta_{\text{R} 4}$& .03 \\
	\hline
	
	Average real variability & ARV & 0.09 \\
	\hline
	
	Power spectrum & PC 1 & < .01 \\
	& PC 2 & < .01 \\
	& PC 3 & .74 \\
	\hline
\end{tabular}}{}
\end{table}
Table~\ref{tabSim2} contains the results for the three approaches under consideration.
As in Section~\ref{frequencySim}, the $p$-values for $z$ were not significant and were left out.
\begin{figure}[tb]
	\centering
	\includegraphics{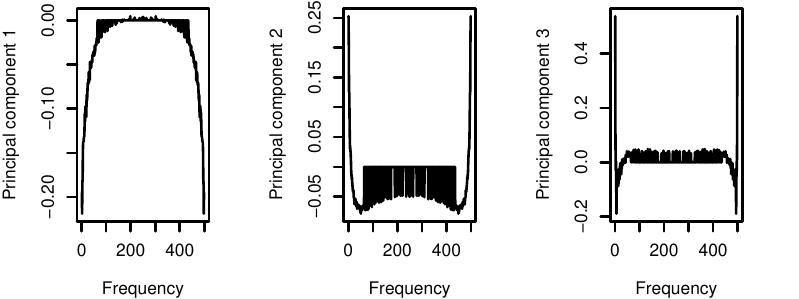}
	\caption{The three functional supervised principal components of the power spectra from the autoregressive simulated data with weighting.}
	\label{sim2PCs}
\end{figure}
Figure~\ref{sim2PCs} contains the supervised principal components used in the power spectrum method.
The weighting functions of the only statistically significant extrema-weighted features $\beta_{\text{L}1}$ and $\beta_{\text{R}4}$ were estimated as $b_{\text{L}1} = .47$ and $b_{\text{
R}4} = .75$, respectively.
The corresponding smooth GAM functions are shown in Figure~\ref{GAMfunctionsAR}.
\begin{figure}[tb]
	\centering
	\includegraphics{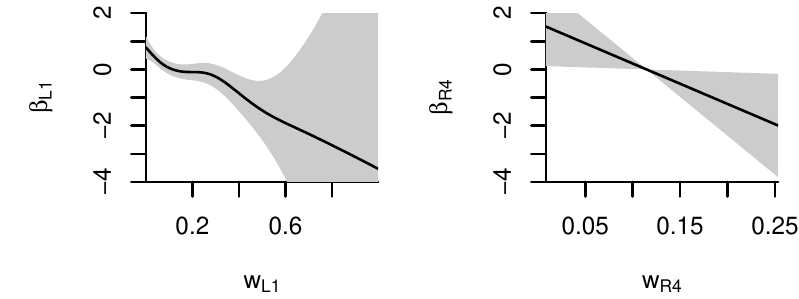}
	\caption{Statistically significant smooth function from the generalized additive model for the autoregressive simulation.}
	\label{GAMfunctionsAR}
\end{figure}

ARV measures the magnitude of $\phi_i$, the persistence, and $v_i$, the marginal variance, jointly. Since the relationship with the outcome is a richer function of $\phi_i$ and $w_i$, ARV fails to find a trend.

The functional supervised principal components are different from the previous simulation due to randomness in the autoregressive innovations. The persistence $\phi_i$ influences the power spectrum, such that this method again detects an association between the functional predictor and the outcome. Still, the functional supervised principal components are hard to interpret to give a meaningful description of the association.

Note that both $\beta_{\text{L} 1}$ and $\beta_{\text{R} 4}$ are estimated as decreasing functions.
The XWFs correctly pick up that it is more lower values, the lower tail, ($w_{\text{L}1}$), and lower derivatives in the right tail ($w_{\text{R}1}$) that correspond to decreased $P(y_i=1)$: A higher persistence $\phi_i$ corresponds with a lower derivative since the marginal variance is fixed in this simulation

The estimated weighting functions with $b_{\text{L}1} = .47$, that is weighting decreases above the 47nd percentile, and $b_{\text{R}4} = .75$ do not match the simulation since $x_i(k)=2$ corresponds to the 92nd percentile. Of note here is that the estimated weighting function is not of the same form as the data generating process and we did not characterize the uncertainty around the estimates of $b_{\text{L}1}$ and $b_{\text{R}4}$.

\subsection{Blood pressure data}

\begin{table}[tb]
\centering
\caption{Results from the blood pressure data\label{tabBP}} {\begin{tabular}{@{}lll@{}}
\hline
Method & Parameter & $p$-value \\
	\hline
	
	Extrame-weighted features & $\beta_{\text{L} 1}$& < .01 \\
	& $\beta_{\text{R} 1}$& .04 \\
	& $\beta_{\text{L} 2}$& .82 \\
	& $\beta_{\text{R} 2}$& < .01 \\
	& $\beta_{\text{L} 3}$& .04 \\
	& $\beta_{\text{R} 3}$& .46 \\
	& $\beta_{\text{L} 4}$& .91 \\
	& $\beta_{\text{R} 4}$& .53 \\
	\hline
	
	Average real variability & ARV & < .01 \\
	\hline
	
	Power spectrum & PC 1 & < .01 \\
	& PC 2 & .45 \\
	& PC 3 & .89 \\
	\hline
\end{tabular}}{}
\end{table}

Table~\ref{tabBP} contains the results for the three methods when applied to the blood pressure data. The $p$-values for the $z$ matrix were omitted, being all less than $.01$.
\begin{figure}[tb]
	\centering
	\includegraphics[width=3.2in]{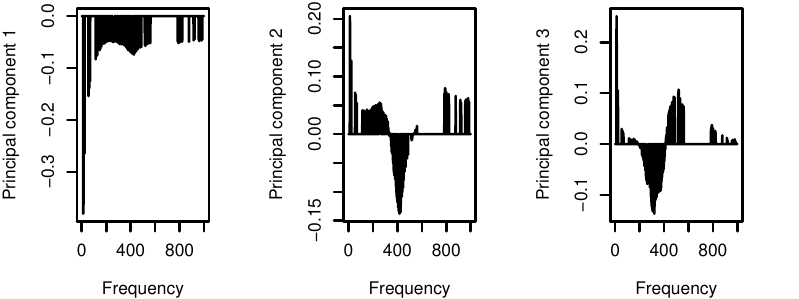}
	\caption{Top five functional principal components of the power spectra from the blood pressure trajectories.}
	\label{BPdataPCs}
\end{figure}

Figure~\ref{BPdataPCs} contains the supervised principal components used in the power spectrum method. For computational stability of the supervised principal components, we truncated the spectra at a frequency of a 1000.
The weighting functions of the statistically significant XWFs were estimated as $b_{\text{L}1} = .03$, $b_{\text{R}1} = .53$, $b_{\text{R}2} = .53$, and $b_{\text{L}3} = .41$.
The corresponding smooth GAM functions are shown in Figure~\ref{GAMfunctionsBP}.
\begin{figure}[tb]
	\centering
	\includegraphics[width=3.2in]{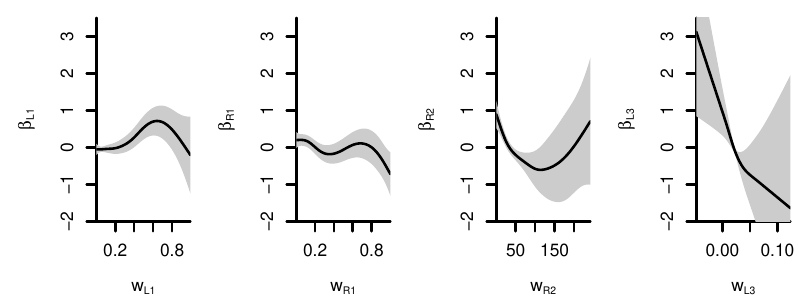}
	\caption{Statistically significant smooth functions from the generalized additive model for the blood pressure application.}
	\label{GAMfunctionsBP}
\end{figure}

\begin{figure*}[tb]
	\centering
	\includegraphics{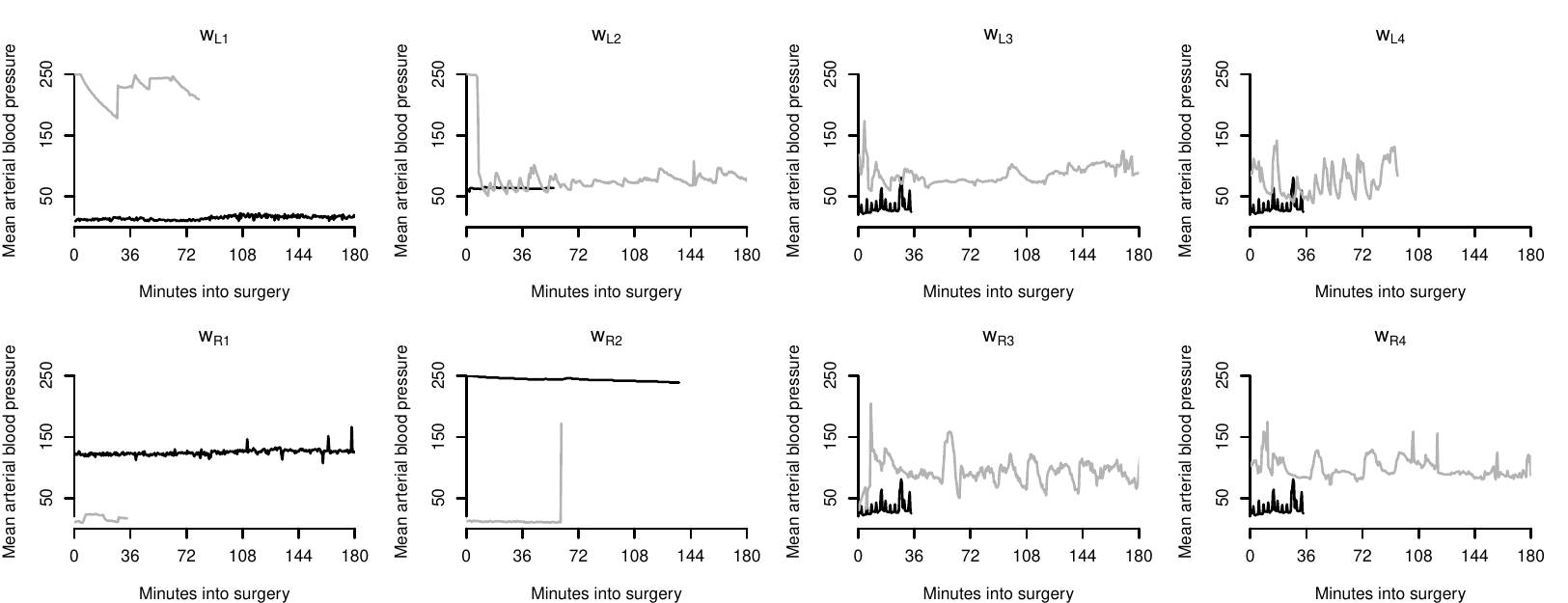}
	\caption{The blood pressure trajectories with the highest (black) and lowest (gray) values for the extrema-weighted features.}
	\label{XWFfig}
\end{figure*}
Figure~\ref{XWFfig} shows the blood pressure trajectories with the most extreme values for the extrema-weighted features, out of all 18,657 trajectories. It is striking how the trajectories vary in length: One of the goals of the XWFs was to be well-defined over and be comparable across functions with greatly varying sizes of domain.

While the XWFs are defined in an interpretable fashion, their values are not always readily gleaned from a trajectory plot. It is for instance hard to eyeball that the black trajectory has a higher XWF value than the gray trajectory for $w_{\text{L}3}$, $w_{\text{L}4}$, $w_{\text{R}3}$, and $w_{\text{R}4}$ in Figure~\ref{XWFfig}.
However, that $w_{\text{L}1}$ and $w_{\text{R}1}$ measure difference in level is clearly visible with the black and gray trajectories clearly separated in terms of average value.

Average real variability and the power spectrum method find an association between the blood pressure trajectories and postoperative mortality. However, only the XWFs reveal both low and high blood pressure values are predictive. When only considering blood pressure level ($w_{\text{L}1}$ and $w_{\text{R}1}$), not change, then the roughly 3\% lowest measurements are most important on the lower end while high blood pressures are indicative in a wider range.

Also, the XWF results indicate that sharp increases in trajectories when blood pressure is low, measured by $\psi_3(x_i, t) = \left( x_i'(t) \right)_+$ in $w_{\text{L}3}$, are a predictive part of blood pressure variability while blood pressure increases, $\psi_4(x_i, t) = \left( x_i'(t) \right)_-$, are not detected as predictive. The ARV and XWF results agree that blood pressure variability is predictive but the XWFs provide what type of variability.

\subsubsection{Predictive performance}

Using specific predefined features instead of general methods that are less constrained in how they capture variability in the data, might hurt the overall explanatory power. To assess whether XWFs suffer from this in the motivating application, we look at predictive performance. We randomly select 1000 cases, 100 with $y_i=1$ and 900 with $y_1=0$, as test data leaving training data of size 17,657. We compute the XWFs, functional supervised principal components, and ARV as before on the training data. Three predictive models are considered: A GAM with $z$ and only XWFs, with XWFs and ARV, and with XWFs and supervised principal components. We compare the performance of these on the test data to assess any loss in explanatory power by being constrained to XWFs.

Area under the curve (AUC) of the receiver operating characteristic quantifies predictive performance. Using 10 repeated random splits of the data into training and test set, we obtain 10 AUCs for each of the three models considered. They range from 0.55 to 0.74 while the differences in AUC between the models range from -0.02 to 0.05, none of which are statistically significant. The usage of the more interpretable predefined XWFs does not seem to lessen explanatory power compared to more traditional approaches which have a greater focus on predictive performance and less on interpretability.

\section{Discussion}

Existing functional feature extraction methods focus on functions that share the same domain and mostly extract global features.
Motivated by rich blood pressure data from surgeries, we introduced a new feature extraction method, called extrema-weighted features, allowing for differing domains while aiming at functional features that are defined in a local nature, like variability.
We compared XWFs to average real variability, a functional feature from the medical literature, and to functional supervised principal components analysis on the power spectrum of the functional predictor.

Simulations reveal that XWFs are not only more easily interpretable due to the way they are defined, but they are also more flexible in detecting associations that do not fit traditional models. This was instrumental in finding a relationship between blood pressure trajectories during surgery and the outcome postoperative mortality. Where the power spectrum principal components detected a hard to interpret association and ARV only detects that variability is predictive, the proposed extrema-weighted feature model found that sharp increases, especially at lower blood pressures, rather than decreases in blood pressure are predictive.
The XWF results directly address questions of interest to the clinician collaborators. The fact that they corroborate some of their conjectures, further supports the merit of XWFs.
Despite their predefined and specific nature, XWFs had the similar explanatory power as competing methods.

This paper presented a set of well-defined features applicable to the data at hand. The ideas behind them and the general framework of allowing varying domains in the functional predictors and taking weighted averages of local phenomena can however be applied in a much larger variety of ways, making the methods from this paper widely applicable to functional data.

Many parts of the XWF model provide choices to adapt the method to other applications. The weighting function can be chosen and parametrized as desired. Similarly, the generalized additive model shows one possibility on how to use the XWFs but the features can be broadly fed into different models and machine learning methods.

We presented a novel functional feature extraction method motivated by modeling the effect of intraoperative blood pressure trajectories on postoperative mortality. The extrema-weighted features proposed are by design suited for functional predictors whose domains vary. The XWFs outperform more traditional methods and functional data analysis tools, both in simulations and in the application of interest. The proposed model constitutes a modifiable framework for interpretable functional feature extraction.

\section*{Acknowledgements}
We would like to thank Dr.~Michael W.\ Manning and Dr.~Tracy L.\ Setji for their discussions of the medical data.

\section*{Funding}

This work has been supported by Accenture PLC.

\bibliographystyle{natbib}
\bibliography{document}

\end{document}